\documentclass[twocolumn,amsmath,amssymb,showpacs]{revtex4}
\usepackage{graphicx}

\begin{document}

\title{Spin excitations in nickelate superconductors}

\author{Tao Zhou$^{1}$}
\email{tzhou@scnu.edu.cn}

\author{Yi Gao$^{2}$}

\author{Z. D. Wang$^{3}$}
\email{zwang@hku.hk}

\affiliation{
$^{1}$Guangdong Provincial Key Laboratory of Quantum Engineering and Quantum Materials, GPETR Center for Quantum Precision Measurement, SPTE,
	and
	Frontier Research Institute for Physics,
	South China Normal University, Guangzhou 510006, China\\
$^{2}$ Department of Physics and Institute of Theoretical Physics,
Nanjing Normal University, Nanjing, 210023, China.\\
$^{3}$Department of Physics and HKU-UCAS Joint Institute for Theoretical and Computational Physics at Hong Kong, The University of Hong Kong, Pokfulam Road, Hong Kong, China.
}

\date{\today}
\begin{abstract}
We study theoretically spin excitations in the newly discovered nickelate superconductors based on a three-band model and the random phase approximation. The spin excitations are found to be incommensurate in a low energy region. The spin resonance phenomenon is revealed as the excitation energy increases.
The maximum intensity may be at the incommensurate momentum or the commensurate momentum, depending on the out-of-plane momentum.
The spin excitations become incommensurate again at higher energies. The similarities and differences of the spin excitations between nickelate
and cuprate superconductors are addressed. Our predicted results can be tested
by inelastic neutron scattering experiments later.
\end{abstract}

\pacs{74.20.Rp, 74.20.Mn, 74.70.-b}

\maketitle

{\bf Keywords}: nickelates, superconductivity, d-wave pairing, spin susceptibility

\section{Introduction}
The high-T$_c$ cuprate superconductivity was discovered over thirty years ago, while a full understanding of its microscopic mechanism has so far been challenging~\cite{palee}. There may be no doubt that exploring the cuprate analogs can help us to discover a new family of superconductors and to understand physics of cuprate superconductors~\cite{yliu,jyang,bhat}. The Ni atom is close to the Cu atom on the periodic table.
The infinite-layer RNiO$_2$ (R=La, Nd) compounds were realized in the early years~\cite{cre,hay,hay1}. The compounds are isostructural to CaCuO$_2$~\cite{sieg}. The latter is a parent compound of cuprate superconductors. The Ni$^{+}$ ions in RNiO$_2$ have a 3d$^{9}$ configuration, similar to the Cu$^{2+}$ ions in the high-T$_c$ cuprate superconductors. Therefore, the nickelates appear to be a most possible candidate material analogous to cuprates, and thus have attracted considerable attention in the past~\cite{cre,hay,hay1,ani,ike,iked,kwlee,liu,bot}.
While the differences between the nickelates and the cuprates have also been reported, namely, the RNiO$_2$ may be metallic rather than a magnetic insulator~\cite{hay,hay1,ike,iked,kwlee,liu,bot}. And at low energies the Ni-d$_{x^2-y^2}$ band may mix the La-5d band for the LaNiO$_2$ compound~\cite{kwlee,liu}.

Very recently, superconductivity with T$_c$ ranging from 9K to 15K was discovered in the Sr doped infinite-layer nickelate material
Nd$_{1-x}$Sr$_{x}$NiO$_2$~\cite{dli}. This new discovery has attracted much attention and many groups have restudied the electronic structure and physical properties of nickelates~\cite{bota,saka,hep,xwu,nomu,jgao,ryee,sing,zhang,jiang,lhhu,hira,bern,guy,qli,fuy,zhoux,sil,lech,chang,liuz,talan}. One of the most crucial issues is to reconsider the differences and similarities between nicklates and cuprates and find the
mechanism to be responsible for the superconductivity in nickelates. Experimentally, different results have also been reported, namely, no superconductivity was observed in the Nd$_{1-x}$Sr$_{x}$NiO$_2$ bulk sample~\cite{qli} or film samples~\cite{zhoux}. In Ref.~\cite{sil}, based on the density functional theory, it was proposed that the 3$d^9$-type RNiO$_2$ is still analogous to the cuprates. The absence of superconductivity in some nickelate samples is due to intersecting Hydrogen, leading to the $3d^{8}$ configuration of Nickel ions.

Although the mechanism of high-T$_c$ superconductivity is still puzzling now, it is generally believed that spin excitations should be important.
Previously, the spin excitations in cuprate superconductors are intensively studied, both experimentally~\cite{ros,tra,moo,mook,bour,pail} and theoretically~\cite{lchen,bulu,bulut,tana,ste,liu1,bulu1,plee,jxli,mans,zhou,zhou1,zhou2}.
Experimentally, the momentum and frequency dependence of spin excitations are studied through the inelastic neutron scattering (INS) experiments.
Theoretically, the spin excitations are explored through the imaginary part of the dynamical spin susceptibility. Generally the experimental observations and theoretical results are qualitatively consistent. In the superconducting state, one of the most important results is the resonant spin excitation at the commensurate momentum $(\pi,\pi)$ and certain resonant energy. Below and above the resonant energy, the intensity of the spin excitation decreases rapidly and the momentum moves to an incommensurate one~\cite{ros,tra,moo,mook,bour,pail,lchen,bulu,bulut,tana,ste,liu1,bulu1,plee,jxli,mans,zhou,zhou1,zhou2}.
For nickelates, the parent compounds RNiO$_2$ are metallic and there is no magnetic order~\cite{hay,hay1,ike,iked,kwlee,liu,bot,dli}. At the first sight, it seems that the nickelates are different from the cuprates. While it has been reported by many groups that the Ni-3d$_{x^2-y^2}$ band is self-doped with holes by La/Nd-5d electrons~\cite{kwlee,liu,bota,saka,hep,xwu,ryee,sing,zhang,jiang,hira}.
Therefore, the disappearance of the magnetic order is understandable. Although no static magnetic order exists for the parent compounds, the dynamical spin fluctuation may still be important for superconductivity.
Recently, it was indeed proposed that the hole-doped nickelate system is magnetically cuprate-like based on the magnetic force theory~\cite{ryee}.
In the mean time, it has been reported that the phonon cannot support such a high superconducting transition temperature for nickelate superconductors~\cite{nomu}.
Therefore, the spin fluctuation may be a good candidate for mediating the electron
pairing.
Actually the possible superconductivity has been explored based on the spin fluctuation theory. The $d$-wave pairing symmetry is proposed~\cite{saka,xwu}.
Therefore, now it is timely and insightful to explore theoretically the spin excitations of nickelates in detail.
The numerical results may be compared with later INS experimental results, which can provide an unambiguous
clue to conclude whether the nickelates are cuprate-like superconductors and help to
seek the mechanism of the superconductivity in this compound.

Several models for the band structure have been proposed recently to describe the nickelates, including the single-band model~\cite{bota,xwu,hira}, two-band model~\cite{hep,jgao,lhhu}, three-band model~\cite{xwu,nomu} or more band model~\cite{saka}.
Generally the spin excitations are determined by the geometry of the Fermi surface~\cite{tana,ste,liu1,plee,zhou}. We here
start from a three-band model and consider the $d$-wave superconducting pairing. The spin excitations are studied based on the random phase approximation (RPA). Our numerical results indicate that quasi-spin-resonance may exist at a typical resonant energy. At low and very high energies the spin excitations are generally incommensurate, which can be  well understood from the scattering of the Fermi surface. The similarities and differences between the spin excitations of nickelates with those of cuprates are discussed.

The rest of the paper is organized as follows.
In Sec. II, we introduce
the model and present the relevant formalism. In Sec. III, we
report numerical calculations and discuss the obtained
results. Finally, we give a brief summary in Sec. IV.

\section{Model and formalism}
We start from the Hamiltonian including the hopping term, the superconducting pairing term, and the interaction term, expressed as,
\begin{equation}
H=H_t+H_p+H_{int}.
\end{equation}

The hopping term $H_t$ is obtained from a three-band tight-binding fit from the first-principal band structure of the Nd$_{0.8}$Sr$_{0.2}$NiO$_2$ compound~\cite{xwu},
\begin{equation}
H_t=-\sum_{{\bf i}{\bf j}\mu\nu\sigma}t^{\mu\nu}_{\bf ij}c^\dagger_{{\bf i}\mu\sigma}c_{{\bf j}\nu\sigma}-\mu_0\sum_{{\bf i}\mu\sigma} c^\dagger_{{\bf i}\mu\sigma}c_{{\bf i}\mu\sigma}.
\end{equation}
Here $c^\dagger_{{\bf i}\mu\sigma}$ ($c_{{\bf i}\mu\sigma}$) are creation (annihilation) operators at the $i$-th site in the orbital
$\mu$ and with spin projection $\sigma$. The indices $\mu$ and $\nu$ run through $1$ to $3$ corresponding to
the Ni-$3d_{x^2-y^2}$ orbital, the Nd-$5d_{3z^2-r^2}$ orbital, and the Nd-$5d_{xy}$ orbital, respectively.

$H_p$ is the superconducting pairing term, expressed as,
\begin{equation}
H_p=\sum_{\bf ij}(\Delta_{\bf ij}c^{\dagger}_{{\bf i}1\uparrow}c^{\dagger}_{{\bf j}1\downarrow}+h.c.).
\end{equation}
Following Ref.~\cite{xwu} and assuming that the electron pairing is mediated by the spin excitations, only the electron pairing within the Ni-$3d_{x^2-y^2}$ orbital is considered.

The electron interaction term $H_{int}$ is expressed as,
\begin{eqnarray}
H_{int}&=&\sum_{{\bf i},\alpha\leq\alpha^\prime} U_{\alpha \alpha^\prime}n_{{\bf i}\alpha\uparrow}n_{{\bf i}\alpha^\prime\downarrow}+J_H\sum_{{\bf i}\sigma\sigma^\prime}c^\dagger_{{\bf i}2\sigma}c^\dagger_{{\bf i}3\sigma^\prime}c_{{\bf i}2\sigma^\prime}c_{{\bf i}3\sigma}\nonumber\\&&
+J^\prime \sum_{\bf i}(c^\dagger_{{\bf i}2\uparrow}c^\dagger_{{\bf i}2\downarrow}c_{{\bf i}3\downarrow}c_{{\bf i}3\uparrow}+h.c.),
\end{eqnarray}
where the $U_{\alpha\alpha}$ and $U_{\alpha\alpha^\prime}$ $(\alpha\neq\alpha^\prime)$ represent the intra-orbital and inter-orbital interactions, respectively. $J_H$ and $J^\prime$ are the Hund's coupling constant and the pairing hopping constant of the two Nd-$5d$ orbitals.

Defining a six order column vector with $\Psi({\bf k})=(c_{{\bf k}1\uparrow},c_{{\bf k}2\uparrow},c_{{\bf k}3\uparrow},c^\dagger_{-{\bf k}1\downarrow},c^\dagger_{-{\bf k}2\downarrow},c^\dagger_{-{\bf k}3\downarrow})^{\mathrm{T}}$,
the above bare Hamiltonian without the interaction term can be expressed as the $6\times 6$ matrix term in the momentum space.
The bare spin susceptibility including both the normal and the anomalous terms is then expressed as~\cite{ygao1},
\begin{eqnarray}
\chi^{\alpha \alpha^\prime}_0({\bf
q},\omega)&=&\frac{1}{N}\sum_{{\bf{k}}ij}[u_{\alpha i}({\bf{k}})u_{\alpha^\prime i}({\bf{k}})u_{\alpha^\prime j}({\bf{k}}+{\bf{q}})u_{\alpha j}({\bf{k}}+{\bf{q}})+\nonumber\\&&u_{\alpha i}({\bf{k}})u_{\alpha^\prime+3, i}({\bf{k}})u_{\alpha^\prime j}({\bf{k}}+{\bf{q}})u_{\alpha+3, j}({\bf{k}}+{\bf{q}})]
\nonumber \\&& \times\frac{f(E_j({\bf{k}}+{\bf{q}}))-f(E_i({\bf{k}}))}{\omega-E_j({\bf{k}}+{\bf{q}})+E_i({\bf{k}})+i\delta},
\end{eqnarray}
where $U_{ij}({\bf{k}})$ and $E_i({\bf{k}})$ are eigenvectors and eigenvalues of the Hamiltonian which can be obtained through diagonalizing
the Hamiltonian matrix. The cases with $\alpha=\alpha^\prime$ and $\alpha\neq \alpha^\prime$ are corresponding to the intra-orbital excitations and inter-orbital excitations, respectively.
Generally both the inter-orbital scattering and the intra-orbital scattering
may contribute to the spin susceptibility.
While from the present band structure, the three orbitals are weakly coupled and the spin susceptibilities from the inter-orbital scattering are nearly vanishing. As a result, only intra-orbital spin excitations (related to intra-orbital interaction terms in Eq.(4)) need to be considered.  Moreover, the renormalized spin susceptibilities contributed by the Nd-layer are negligibly small in the whole momentum and energy space we considered. More details and the numerical verifications have been presented in the supplemental material.
Then, following Refs.~\cite{zhou1,zhou2}, here only the scattering within the Ni-$3d_{x^2-y^2}$ orbital is considered with
$\chi_0({\bf q},\omega)=\chi^{11}_0({\bf q},\omega)$.

The renormalized spin susceptibility can be obtained through the RPA~\cite{lchen,bulut}, which is given by
\begin{equation}
\chi({\bf q},\omega)=\frac{\chi_0({\bf q},\omega)}{1-U \chi_0({\bf q},\omega)}.
\end{equation}
Here $U=U_{11}$ is the on-site Hubbard-like interaction in the Ni-$3d_{x^2-y^2}$ band.

In the following presented results,
we use the nearest-neighbor hopping constant of the Ni-${3d}$ band as the energy unit. Other
in-plane hopping constants are taken from Ref.~\cite{xwu} and presented in the supplemental material.

\section{Results and discussion}

  \begin{figure}
\centering
  \includegraphics[width=2.4in]{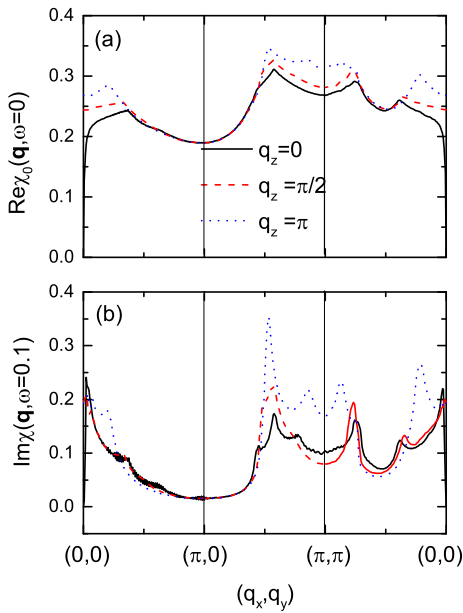}
\caption{(Color online) (a)The real parts of the bare normal state spin susceptibility as a function of the momentum ${\bf q}$ with $\omega=0$.
(b) The imaginary parts of the renormalized normal state spin susceptibility as a function of the momentum ${\bf q}$ with $\omega=0.1$.
}
\end{figure}

We now study the spin excitations in the normal state by setting the gap magnitude $\Delta_0=0$ in Eq.~(1).
 At first we shall justify the effective on-site interaction $U$ in the RPA factor. In the RPA theory, the effective interaction $U$ differs from the original on-site repulsive interaction. Usually the former is much smaller than the latter to ensure the RPA framework to be correct~\cite{lchen,bulut}. Also, for the $t-J$ type model, it was proposed that the renormalized interaction
in the RPA theory should multiply an additional factor $\alpha=0.34$ to match the antiferromagnetic instability~\cite{plee}. Therefore, it is difficult to determine $U$ directly from the band calculation. While one can obtain an effective range for the value of $U$ from the magnetic instability. In the framework of RPA, the magnetic instability occurs when the pole condition of the real part of the zero energy RPA factor occurs [$1-U\mathrm{Re}\chi_0({\bf q},0)=0$].
We plot the real part of the zero energy normal state bare spin susceptibility as a function of the in-plane momentum ${\bf q}$ in Fig.~1(a). As is seen, the maximum value of Re$\chi_0({\bf q},0)$ is about 0.346. To avoid the magnetic instability, the effective $U$ should be taken as $U<2.89$. Note that the doping density is 0.2, which is far away from the antiferromagnetic instability point.  In the following presented results, we choose $U=2$. We have checked numerically that the results are stable when the value of $U$ changes slightly.

The imaginary parts of the spin susceptibility as a function of the in-plane momentum ${\bf q}$ with the energy $\omega=0.1$ are plotted in Fig.~1(b). As is seen, the main contributions of the spin excitations are around the antiferromagnetic momentum $(\pi,\pi)$.
The maximum spin excitations occur at an incommensurate in-plane momentum ${\bf Q_{\parallel}}$ with ${\bf Q_{\parallel}}\approx(0.55\pi,\pi)$.  The maximum intensity increases as the out-of-plane momentum $q_z$ increases. The momentum ${\bf Q_{\parallel}}$ depends weakly on $q_z$.

  \begin{figure}
\centering
  \includegraphics[width=2.4in]{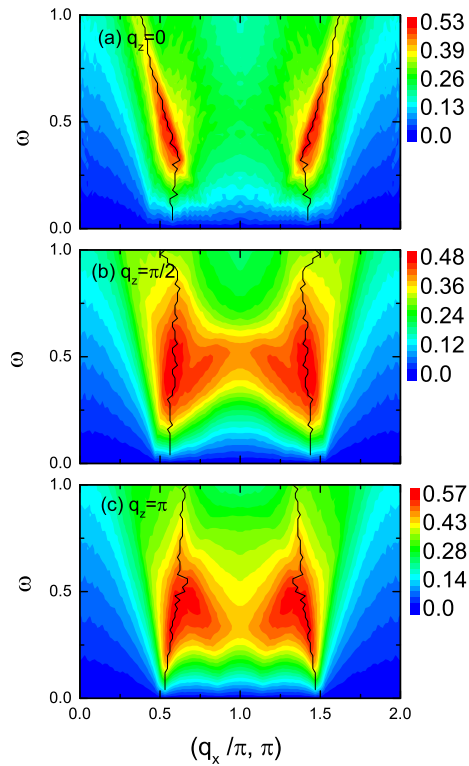}
\caption{(Color online) The intensity plots for the imaginary parts of the spin
susceptibilities in the normal state as functions of the momentum (with $q_y=\pi$) and the energy.
}
\end{figure}

The intensity plots of the imaginary parts of the spin susceptibilities
as functions of the momentum and the energy in
the normal state with different $q_z$ are plotted in Figs. 2(a)-2(c).
For the case of $q_z=0$, the main spin excitations are around the incommensurate in-plane momentums $(\pi\pm\delta,\pi)$.
$\delta$ is defined as the incommensurability. As $q_z$ increases, the spin excitations near the in-plane momentum $(\pi,\pi)$ increases while the maximum
spin excitations are still at an incommensurate momentum.

  \begin{figure}
\centering
  \includegraphics[width=2.4in]{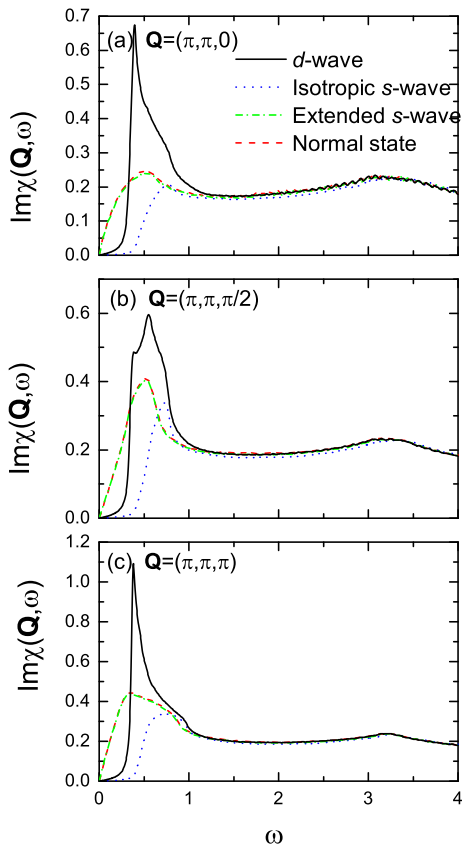}
\caption{(Color online) The imaginary parts of the spin
susceptibilities as a function of the energy.
}
\end{figure}

We turn to study the spin excitations in the superconducting state.
At first, we would like to discuss the possible superconducting pairing symmetry. The nickelate compound crystallizes in a tetragonal structure. As indicated in Ref.~\cite{dli}, the Nd$_{1-x}$Sr$_{x}$NiO$_2$ material is in the
infinite-layer phase,
with NiO$_2$ planes
intersecting with the Sr-doped Nd-layers. Due to the existence of the Nd-layers, the coupling between different NiO$_2$ layers should be weak.
This is also consistent with the tight-binding band parameters from the band calculation~\cite{xwu}, namely, the inter-layer hopping constants for the Ni-$3d_{x^2-y^2}$ orbital are much smaller that the intra-layer nearest-neighbor hopping constants.
Therefore, only the electron pairings within the NiO$_2$ plane are considered in the present work.
There are three possible pairing symmetries. For the case of ${\bf j}={\bf i}$ in Eq.(3),
 the pairing symmetry is
 the isotropic $s$-wave pairing with $\Delta_{\bf k}\equiv \Delta_0$.
 When ${\bf j}$ is the nearest-neighbor site to ${\bf i}$,
the pairing symmetry may be the extended $s$-wave pairing state with $\Delta_{\bf ij}\equiv \Delta_0/4$,
or the $d_{x^2-y^2}$-pairing state with $\Delta_{\bf ij}=\pm \Delta_0/4$ ($\pm$ depend on the bond $\langle ij\rangle$ being along the $x$ direction or the $y$ direction). In the momentum space, through Fourier transformation, these two pairing symmetries are expressed as $\Delta_{\bf k}=\Delta_0(\cos k_x+\cos k_y)/2$ and
$\Delta_{\bf k}=\Delta_0(\cos k_x-\cos k_y)/2$, respectively.

The most possible pairing symmetry can be further investigated from the above normal state spin excitations presented in Fig.~1, namely,
the NiO$_2$ planes are hole-doped from the half-filled Mott insulators. There may exist significant spin excitations
near the antiferromagnetic in-plane momentum $(\pi,\pi)$, which may in principle generate in-plane superconducting $d$-wave pairing.
Previously based on the spin fluctuation scenario, the $d_{x^2-y^2}$ pairing symmetry is indeed supported~\cite{saka,xwu}.
In the following, we follow Refs.~\cite{saka,xwu} and consider the pairing symmetry as the $d$-wave pairing within the NiO$_2$ plane. For the infinite-layer phase, each NiO$_2$ plane is equivalent. Therefore, the gap magnitude $\Delta_{0}=0.2$ is considered below, independent on the $z$-axis coordinates.

Now let us study the energy dependence of the spin excitations.
One of most important results in the superconducting state is the resonant spin excitations. The possible spin resonance has been studied theoretically in many unconventional superconducting systems,
including the cuprates~\cite{bulu1,plee,jxli,mans,zhou,zhou1,zhou2}, the iron-based superconductors~\cite{kuro,maie,kemp,ygao,ygao1}, the Na$_x$CoO$_2\cdot$yH$_2$O superconductors~\cite{jianxin}, and the heavy fermion material CeCoIn$_5$~\cite{chub}.
Here we would like to explore whether the spin resonance exists in nickelate superconductors. The imaginary parts of the spin susceptibilities as a function of the energy at the in-plane momentum ${\bf Q_{\parallel}}=(\pi,\pi)$ are displayed in Fig.~3. For comparison, we also present the spin susceptibilities in the normal state, the isotropic $s$-wave pairing state and the extended $s$-wave pairing state in Fig.~3. Then it is seen clearly that a spin resonance peak occurs near the energy $0.4$ [about twice of the gap magnitude] for the spin excitation in the $d$-wave superconducting state. For the isotropic $s$-wave pairing state, the spin excitations are nearly vanishing within the superconducting gap and no spin resonance occurs.
For the extended $s$-wave pairing state, no gap feature is seen. The spin excitation is almost the same as that of the normal state and no spin resonance exits, either. This can be explained well through exploring the nodal lines of the extended $s$-wave pairing.
They coincide with the boundaries of the magnetic Brillouin zone, expressed as $(k_x=\pi\pm k_y)$.
The spin excitations at the momentum $(\pi,\pi)$ are mainly contributed by the scattering between boundaries of the magnetic Brillouin zone.
Thus the superconducting pairing has no effect on the $(\pi,\pi)$ spin excitation for this extended $s$-wave pairing symmetry.

The energy dependence of the spin excitations may be studied by later INS experiments. The existence of the spin resonance at the in-plane momentum $(\pi,\pi)$ may be tested and taken as one important signature for the $d$-wave superconductivity.
The spin resonance is robust for different $q_z$. The position changes slightly as $q_z$ changes.
  \begin{figure}
\centering
  \includegraphics[width=2.8in]{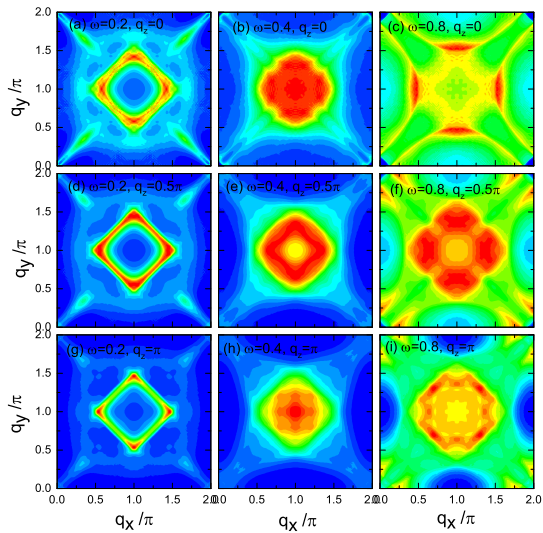}
\caption{(Color online) The intensity plots of the imaginary
part of the spin susceptibility as a function of the in-plane momentum in the superconducting state with different energies and different out-of-plane momentum $q_z$.
}
\end{figure}

The intensity plots of the imaginary part of the spin susceptibility as a function of the in-plane momentum with different energies and $q_z$ are presented in Fig.~4.
For the case of $q_z=0$, when the energy is below the resonant energy, as is seen in Fig.~4(a), four incommensurate peaks at the momentums $(\pi\pm\delta,\pi)$ and $(\pi,\pi\pm\delta)$ are seen clearly. As the energy increases to the resonant energy, as is seen in Fig.~4(b), the spin excitation at the momentum $(\pi,\pi)$ increases greatly, while the spin excitation is still incommensurate at this energy and the incommensurability is rather small. When the energy increases further to above the resonant energy [Fig.~4(c)], the spin excitation is still incommensurate while the incommensurability is large. The numerical results for $q_z=0.5\pi$ are displayed in Figs.~4(d)-4(f). At low energy with $\omega=0.2$, the spin excitation is also incommensurate. The maximum spin excitations form a circle around $(\pi,\pi)$. Near the resonant energy, the maximum spin excitation still occurs at an incommensurate momentum and the incommensurability is still rather large. Above the resonant energy with $\omega=0.8$, the spin excitation is also incommensurate.
When the out-of-plane momentum $q_z$ increases to $\pi$, as is seen in Figs.~4(g)-4(i), the spin excitations are incommensurate at low and high energies. Near the resonant energy, the spin excitation is commensurate.

  \begin{figure}
\centering
  \includegraphics[width=2.4in]{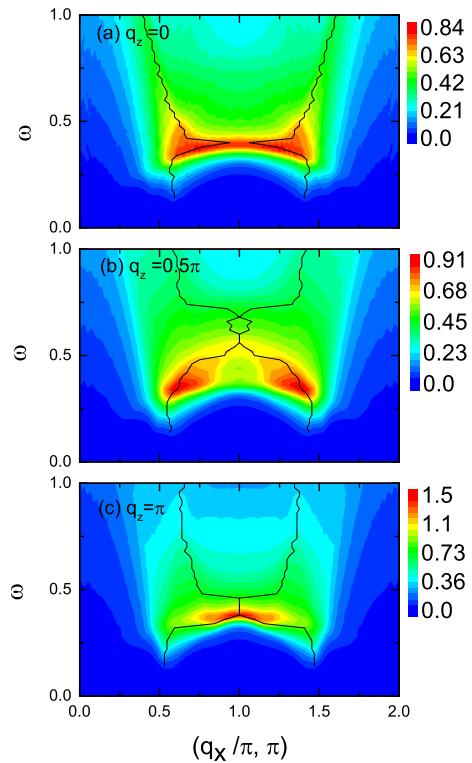}
\caption{(Color online) The intensity plots of the imaginary
part of the spin susceptibility as a function of the in-plane momentum (with $q_y=\pi$) and the energy in the superconducting state with different out-of-plane momentum $q_z$.
}
\end{figure}

To study the energy dependence of the spin excitations more clearly, the intensity plots of the imaginary part of the spin susceptibility as functions of the momentum and energy are presented in Fig.~5.
The incommensurability reaches the minimum at the energy about 0.4. For higher energies ($\omega>0.6$), the superconducting order parameter plays a relatively minor role and the dispersion at this energy region is similar to those of the normal state [presented in Fig.~2]. As $q_z$ increases to 0.5, the spin excitation is commensurate near the energy $0.6$. While at the energy 0.4, where the spin resonance is excepted, the spin excitation is still
incommensurate. Actually, the spin excitations at this momentum are similar to those in La-based cuprate superconductors~\cite{zhou,vign}. As discussed in Ref.~\cite{zhou}, when the spin resonance is weak, the incommensurate spin excitation may occur. The commensurate spin excitation at higher energies is consistent with the bare spin susceptibility.
For the case of $q_z=\pi$, as is seen, the dispersion is downward at low energies and upward at high energies. The spin excitation is commensurate near the resonant energy.
For all of $q_z$ we considered, a hourglass dispersion is seen clearly.
Note that, the dispersion may also be obtained from INS experiments.
For cuprate superconductors, the hourglass dispersion has also been reported both experimentally and theoretically~\cite{bour,pail,plee,zhou}.
Actually, the dispersion of the maximum spin excitations is closely related to the band structure, the Fermi surface, and the pairing symmetry of the material.
We expect that our theoretical predictions for the dispersion may be tested by later INS experiments. Then a lot of useful information may be provided.

Generally the numerical results of spin excitations can be well understood based on the geometry of Fermi surface. We plot the normal state Fermi surface with different $k_z$ in Fig.~6(a).
As is seen, as $k_z=0$, the normal state Fermi surface including a small pocket around the $(0,0)$ point is contributed mainly by the Nd-5d$_{3z^2-r^2}$ orbital and a large Fermi surface is mainly contributed by the Ni-3d$_{x^2-y^2}$ band. The small Fermi pocket disappears completely as $k_z$ increases to larger than $0.21\pi$. As $k_z$ increases further, when $k_z>0.81\pi$, another small Fermi pocket around the $(\pi,\pi)$ pocket contributed mainly by the Nd-5d$_{xy}$ orbital appears.
Based on the first principle calculation, the inter-orbital hopping constants are much smaller than the intra-orbital hopping ones~\cite{xwu}. Therefore, here the spin excitations are mainly determined by the scattering between the same Fermi pocket. The large Fermi surface is close to the magnetic Brillouin zone, generating significant spin excitations around the momentum $(\pi,\pi)$, as indicated by the solid vectors ${\bf Q_1}$ in Fig.~6(a).
The low energy spin susceptibility comes from the
particle-hole excitations around the Fermi surface. Generally, when the Fermi pocket is small, the low energy spin excitations are also small. Therefore, here the spin susceptibility contributed by the Nd-layer is generally much smaller than that by the Ni-$3d_{x^2-y^2}$ band.
As a result, the scattering between the small Fermi pocket from Nd-5d band
will merely enhance the spin excitation at the small momentums while it is not important to the superconductivity.
On the other hand, now the minimum model for nickelates is still under debate. It was proposed that the Nd-$5d$ orbitals may be strongly coupled to the Ni-3d$_{x^2-y^2}$ orbital to form the Kondo spin singlets~\cite{zhang}. In this case,
there may exist some additional peaks of the spin susceptibilities
contributed by the inter-orbital scattering, which can be discussed from the normal state Fermi surface shown in Fig.~6(a). As is seen, the dashed arrows indicate the vectors connecting the Ni-3d$_{x^2-y^2}$ orbital and the Nd-$5d$ orbitals, with the vectors ${\bf Q_{2,3}}=(\pi/2\pm \delta,\pi/2\pm \delta)$ $(\delta=0.1\pi)$. Based on the Fermi surface nesting picture~\cite{plee}, the inter-orbital coupling may generate the spin excitations around the momentum $(\pi/2,\pi/2,q_z)$. Therefore, if the Nd-$5d$ orbitals and the Ni-3d$_{x^2-y^2}$ orbital are indeed strongly coupled, the spin excitations around the momentum $(\pi,\pi,q_z)$ (from the intra-orbital scattering) and the spin excitations around the momentum $(\pi/2,\pi/2,q_z)$ (from the inter-orbital scattering) may coexist in the system. Thus the spin excitations may also provide a potential justification of the models if such coexistence is detected by experiments.

For the case of low energy spin susceptibility in the superconducting state, only excitations near nodes can occur. Thus qualitatively speaking, the low energy spin excitations are qualitatively similar for different $q_z$. A more insightful explanation for the low energy incommensurate spin excitations can be obtained through the constant energy contours.
According to Ref.~\cite{plee}, at low energies, the spin excitations are mainly determined by the bare spin susceptibilities. Generally,
the scattering between the energy contours $E_{\bf k}=\omega_j/2$ is responsible for the
spin excitations at the energy $\omega_j$. The contour plots of the quasiparticle energy with $E_{\bf k}=0.1$ for different $k_z$ are plotted in Fig.~6(b). As is seen, here for all of $k_z$ we considered, the contours for different $k_z$ are qualitatively similar. Therefore, the spin excitations at the energy $\omega=0.2$ are qualitatively similar and are incommensurate for different our-of-momentum $q_z$.

  \begin{figure}
\centering
  \includegraphics[width=2.4in]{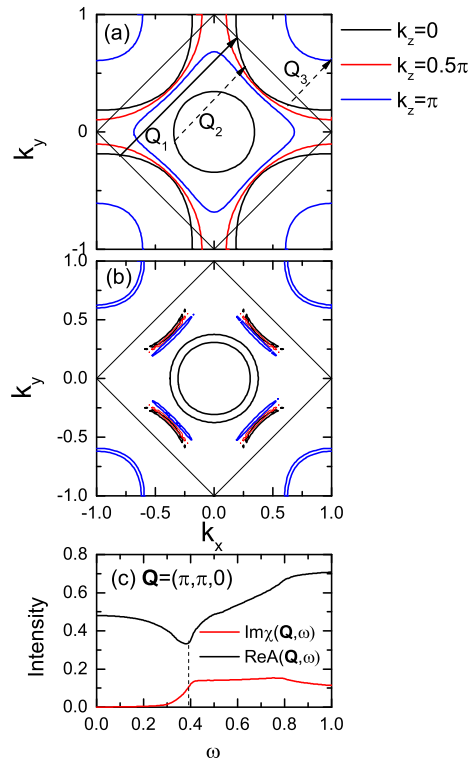}
\caption{(Color online) (a) The normal state Fermi surfaces. The solid arrow indicates the intra-orbital scattering with ${\bf Q_1}=(\pi,\pi)$. The two dashed arrows indicate the inter-orbital scattering with
${\bf Q_2}=(0.6\pi,0.6\pi)$ and ${\bf Q_3}=(0.4\pi,0.4\pi)$. (b) The constant energy contours. (c) The real part of the RPA factor and the imaginary part of the bare spin susceptibility as a function of the energy.
}
\end{figure}

We depict the real part of the RPA factor Re$A({\bf q},\omega)$ ($A=1-U\chi_0$) and the imaginary part of the bare spin susceptibility Im$\chi_0$ in Fig.~6(c) to look into the mechanism of the spin resonance.
Here the spin resonance arises from the RPA renormalized effect.
Firstly let us summarize the mechanism for the spin resonance in cuprate superconductors~\cite{plee,jxli}.
At low energies, the energy contours do not touch the hot spot (the crossing points
of Fermi surface with the magnetic Brillouin zone boundary).
Therefore, Im$\chi_0$ tends to zero at the in-plane momentum $(\pi,\pi)$, i.e., a spin gap exists. The spin gap closes when the constant energy contour reaches the hot spot. Usually for hole-doped cuprate compounds, the hot-spot is near the momentum $(\pi,0)$. Thus the spin gap is about $2\Delta_0$ [$\Delta_0$ is the gap magnitude at the momentum $(\pi,0)$].
Due to the flat band at this momentum (extended van Hove
singularity), a step-like rise of Im$\chi_0$ occurs at the edge of the spin-gap. Then a logarithmic singularity in Re$\chi_0$ occurs via the
Kramers-Kronig relation. This singularity will lead to the pole condition of the RPA factor being satisfied within the spin-gap. As a result, a sharp resonance peak appears for the imaginary part of the renormalized spin susceptibility. For nickelates compound, the nodal point at the Fermi surface is away from the magnetic Brillouin zone for all of $k_z$, as is seen in Fig.~4(a). Thus the spin-gap still exists. While the hot spot depends on $k_z$. As $k_z$ equals to zero, the hot spot is far away from $(\pi,0)$. Thus the real spin gap is less than $2\Delta_0$, as is seen in Fig.~6(c).
Im$\chi_0$ increases gradually at the energy about $0.2$.
Therefore, there is no step-like rise for Im$\chi_0$. At the energy $2\Delta_0$, there is still a peak for Re$\chi_0$ due to the flat band dispersion at $(\pi,0)$, although there is no singularity at this energy. Then ReA($\omega$) reaches the minimum at this energy, leading to the quasi-resonance behavior. Here the spin excitation is indeed enhanced for a typical $d$-wave superconducting state. As the pole condition is not really satisfied, the RPA renormalized effect is not as strong as that in cuprates.
Therefore, for some $q_z$ the spin excitation may still be incommensurate even at the resonant energy.

As the energy increases further, the RPA renormalization has a rather weak effect. Then the spin excitations are still determined by the bare spin susceptibility, similar to the case of low energy excitations. While at the high energy, the antinodal to antinodal excitations play an important role.
The normal state Fermi surface splits at the antinodal direction for different $k_z$. Thus the spin excitations for $q_z=0$ and $q_z\neq 0$ are different. For the case of $q_z=0$, only the scattering between the same Fermi surface occurs. For the case of $q_z\neq 0$, the spin excitations are determined by scattering between the Fermi surface with different $k_z$. Then the weak disorder is induced. It is noted that the spin excitations along different directions are almost the same as those shown in Fig.~4.

Let us summarize the similarities and differences of the spin excitations in the $d$-wave superconducting state between the cuprates and nickelates.
Firstly, although the band structure of nickelates is three dimensional, the properties of the spin excitations are qualitatively similar for different $q_z$.
The maximum spin excitations are always around the momentum $(\pi,\pi)$, which is indeed similar to those of cuprates.
Secondly, in both compounds, a spin resonance phenomena can be observed, namely, the spin excitation is enhanced greatly in the superconducting state at the energy about $2\Delta_0$. The third similarity is that the spin excitations are incommensurate at low and high energies.
Especially, the low energy spin excitations are nearly identical to those of cuprates. On the other hand, there still exist some differences for the spin excitations of these two families. In nickelates, the spin resonance is damped greatly. Therefore for some $q_z$, the maximum excitation does no occur at the momentum $(\pi,\pi)$.
Moreover, the incommensurate spin excitations depend on the out-of-plane momentum $q_z$. The maximum points may form a circle at the momentum space for certain $q_z$. These similarities and differences may be detected by later INS experiments.

At last, we would also like to discuss the difference of spin excitations between nickelates and iron-based superconductors.
In iron-based superconductors, as discussed in Ref.~\cite{kuro}, the $d$-orbitals from iron ions are heavily entangled thus rather strong hybridization exists. Moreover, each Fermi pocket is contributed by several
 orbitals.
 Therefore, the inter-pocket scattering is important to determine their spin excitations~\cite{kuro,maie,kemp,ygao,ygao1}.
For nickalates, although a three-orbital model is taken to generate three Fermi surface pockets,
while from the present band structure we considered, each Fermi pocket is mainly contributed by one orbital and the inter-pocket coupling is weak. As have discussed, the spin excitations are mainly contributed by the Ni-3d$_{x^2-y^2}$ band. The
other two Nd-5d orbitals play a minor role. Also, based on our present model the spin excitations from the inter-pocket scattering are also rather small.

\section{summary}

In summary, starting from a three-band model and $d$-wave superconductivity, we have examined the spin excitations in the nickelate superconductors based on the random phase approximation. A spin resonance phenomenon, namely, the spin excitation is enhanced in the superconducting state at the energy about twice of the gap magnitude, was revealed.
Below and above the resonant energy, the spin excitations are incommensurate. The similarities and differences for the spin excitations in the nickelates and cuprates have been discussed. All of the numerical results are explained well based on the geometry of Fermi surface.

\begin{acknowledgments}
	 This work was supported by the NKRDP of China (Grant No. 2016YFA0301800), and the GRF (Grants No. HKU 173309/16P and No. HKU173057/17P) and a CRF (No. C6005-17G) of Hong Kong.
\end{acknowledgments}


\begin{thebibliography}{References}

\bibitem{palee} P. A. Lee, N. Nagaosa, and X. G. Wen, Rev. Mod. Phys. {\bf 78}, 17 (2006).
\bibitem{yliu} Y. B. Liu, Y. Liu, W. H. Jiao, Z. Ren, and G. H. Cao, Sci. China Phys. Mech. Astron. {\bf 61}, 127405 (2018).  
\bibitem{jyang} J. Yang, T. Oka, Z. Li, H. X. Yang, J. Q. Li, G. F. Chen, and G. Q. Zheng, Sci. China Phys. Mech. Astron. {\bf 61}, 117411 (2018).
\bibitem{bhat} A. Bhattacharyya, D. T. Adroja, M. Smidman, and V. K. Anand. Sci. China Phys. Mech. Astron. {\bf 61}, 127402 (2018).  
\bibitem{cre} M. Crespin, P. Levitz and L. Gatineau, J. Chem. Soc.,
Faraday Trans. 2 {\bf 79}, 1181 (1983).
\bibitem{hay} M. A. Hayward, M. A. Green, M. J. Rosseinsky and J.
Sloan, J. Amer. Chem. Soc. {\bf 121}, 8843 (1999).
\bibitem{hay1} M. A. Hayward and M. J. Rosseinsky, Solid State Sci. {\bf 5}, 839
(2003).
\bibitem{sieg} T. Siegrist, S. M. Zahurak, D. W. Murphy, and R. S. Roth, Nature {\bf 334}, 231 (1998).
\bibitem{ani} V. I. Anisimov, D. Bukhvalov, and T. M. Rice, Phys. Rev. B {\bf 59}, 7901 (1999).

\bibitem{ike} A. Ikeda, T. Manabe and M. Naito, Physica C {\bf 495}, 134
(2013).
\bibitem{iked} A. Ikeda, Y. Krockenberger, H. Irie, M. Naito and H.
Yamamoto, Appl. Phys. Express {\bf 9}, 061101 (2016).
\bibitem{kwlee} K. W. Lee and W. E. Pickett, Phys. Rev. B {\bf 70}, 165109 (2004).
\bibitem{liu} T. Liu, H. Wu, T. Jia, X. Zhang, Z. Zeng, H. Q. Lin and
X. G. Li, AIP Advances {\bf 4}, 047132 (2014).
\bibitem{bot} A. S. Botana and M. R. Norman, Phys. Rev. Materials
{\bf 2}, 104803 (2018).

\bibitem{dli} D. Li, K. Lee, B. Y. Wang, M. Osada, S. Crossley, H. R. Lee, Y. Cui,
Y. Hikita, and H. Y. Hwang, Nature {\bf 572},
624 (2019).

\bibitem{bota} A. S. Botana and M. R. Norman, Phys. Rev. X {\bf 10}, 011024 (2020).
\bibitem{saka} H. Sakakibara, H. Usui, K. Suzuki, T. Kotani, H. Aoki, and K. Kuroki, arXiv: 1909.00060.
\bibitem{hep} M. Hepting, D. Li, C. J. Jia, H. Lu, E. Paris, Y. Tseng, X. Feng, M. Osada, E. Been, Y. Hikita, Y.-D. Chuang, Z. Hussain, K. J. Zhou, A. Nag, M. Garcia-Fernandez, M. Rossi, H. Y. Huang, D. J. Huang, Z. X. Shen, T. Schmitt, H. Y. Hwang, B. Moritz, J. Zaanen, T. P. Devereaux, and W. S. Lee, Nat. Mater. DOI: 10.1038/s41563-019-0585-z (2020).
\bibitem{xwu} X. Wu, D. D. Sante, T. Schwemmer, W.
Hanke, H. Y. Hwang, S. Raghu, and R. Thomale, Phys. Rev. B {\bf 101}, 060504 (2020).
\bibitem{nomu} Y. Nomura, M. Hirayama, T. Tadano,
Y. Yoshimoto, K. Nakamura, and R. Arita, Phys. Rev. B {\bf 100}, 205138 (2019).
\bibitem{jgao} J. Gao, Z. Wang, C. Fang, and H. Weng, arXiv: 1909.04657.
\bibitem{ryee} S. Ryee, H. Yoon, T. J. Kim, M. Y. Jeong, and M. J. Han, Phys. Rev. B {\bf 101}, 064513 (2020).
\bibitem{sing} N. Singh, arXiv: 1909.07688.
\bibitem{zhang} G. M. Zhang, Y. F. Yang, and F. C. Zhang, Phys. Rev. B {\bf 101}, 020501(R) (2020).
\bibitem{jiang} P. Jiang, L. Si, Z. Liao, and Z. Zhong, Phys. Rev. B {\bf 100}, 201106 (2019).
\bibitem{lhhu} L. H. Hu and C. Wu, Phys. Rev. Research {\bf 1}, 032046 (2019).
\bibitem{hira} M. Hirayama, T. Tadano, Y. Nomura, and R. Arita, Phys. Rev. B {\bf 101}, 075107 (2020).
\bibitem{bern} F. Bernardini, V. Olevano, and A. Cano, Phys. Rev. Research {\bf 2}, 013219 (2020).

\bibitem{guy} Y. Gu, S. Zhu, X. Wang, J. Hu, and H. Chen, arXiv: 1911.00814.
\bibitem{qli} Q. Li, C. He, J. Si, X. Zhu, Y. Zhang, and H. H. Wen, arXiv: 1911.02420.
\bibitem{fuy} Y. Fu, L. Wang, H. Cheng, S. Pei, X. Zhou, J. Chen, S. Wang, R. Zhao, W. Jiang, C. Liu, M. Huang, X. Wang, Y. Zhao, D. Yu, F. Ye, S. Wang, and J. W. Mei, arXiv: 1911.03177.

\bibitem{zhoux} X. R. Zhou, Z. X. Feng, P. X. Qin, H. Yan, S. Hu, H. X. Guo, X. N. Wang, H. J. Wu, X. Zhang, H. Y. Chen, X. P. Qiu, and Z. Q. Liu, Rare Metals {\bf 39}, 368 (2020).
\bibitem{sil} Liang Si, Wen Xiao, Josef Kaufmann, Jan M.Tomczak, Yi Lu, Zhicheng Zhong, and Karsten Held, arXiv: 1911.06917.


\bibitem{lech} F. Lechermann, Phys. Rev. B {\bf 101}, 081110 (2020).
\bibitem{chang} J. Chang, J. Zhao, and Y. Ding, arXiv: 1911.12731.
\bibitem{liuz} Z. Liu, Z. Ren, W. Zhu, Z. F. Wang, and J. Yang, arXiv: 1912.01332.
\bibitem{talan} E. F. Talantsev, Results in Physics {\bf 17}, 103118 (2020).

\bibitem{ros} J. Rossat-Mignod, L. P. Regnault, C. Vettier, P. Bourges, P. Burlet,
J. Bossy, J. Y. Henry, and G. Lapertot, Physica C {\bf 86}, 185 (1991).

\bibitem{tra} J. M. Tranquada, P. M. Gehring, G. Shirane, S. Shamoto,
and M. Sato, Phys. Rev. B {\bf 46}, 5561 (1992).

\bibitem{moo} H. A. Mook, Pengcheng Dai, S. M. Hayden, G. Aeppli,
T. G. Perring, and F. Do$\mathrm{\breve{g}}$an, Nature {\bf 395}, 580, 1995.


\bibitem{mook} S. M. Hayden, H. A. Mook, P. Dai, T. G. Perring, and F. Do$\mathrm{\breve{g}}$an, Nature {\bf 429}, 531 (2004).

\bibitem{bour} P. Bourges, Y. Sidis, H. F. Fong, L. P. Regnault, J. Bossy,
A. Ivanov, and B. Keimer, Science {\bf 288}, 1234 (2000).
\bibitem{pail} S. Pailh$\mathrm{\grave{e}}$s, Y. Sidis, P. Bourges, V. Hinkov, A. Ivanov, C. Ulrich, L. P. Regnault, and B. Keimer, Phys. Rev. Lett. {\bf 93}, 167001 (2004).

\bibitem{lchen} L. Chen, C. Bourbonnais, T. Li, and A.-M. S. Tremblay, Phys. Rev. Lett. {\bf 66}, 369 (1991).
\bibitem{bulu} N. Bulut and D. J. Scalapino, Phys. Rev. B {\bf 47}, 3419 (1993).
\bibitem{bulut} N. Bulut, D. J. Scalapino, and S. R. White, Phys. Rev. B {\bf 47}, 2742 (1993).


\bibitem{tana} T. Tanamoto, H. Kohno, and H. Fukuyama, J. Phys. Soc. Jpn. {\bf 63}, 2739 (1994).

\bibitem{ste} G. Stemmann, C. P$\mathrm{\acute{e}}$pin, and M. Lavagna, Phys. Rev. B {\bf 50}, 4075 (1994).

\bibitem{liu1} D. Z. Liu, Y. Zha, and K. Levin, Phys. Rev. Lett. {\bf 75}, 4130 (1995).
\bibitem{bulu1}  N. Bulut and D. J. Scalapino, Phys. Rev. B {\bf 53}, 5149 (1995).
\bibitem{plee} J. Brinckmann and P. A. Lee, Phys. Rev. Lett. {\bf 82}, 2915 (1999).

\bibitem{jxli} J. X. Li, C. Y. Mou, and T. K. Lee, Phys. Rev. B {\bf 62}, 640 (2000).
\bibitem{mans} D. Manske, I. Eremin, and K. H. Bennemann, Phys. Rev. B {\bf 63}, 054517 (2001).
\bibitem{zhou} T. Zhou and Z. D. Wang, Phys. Rev. B {\bf 76}, 094510 (2007).
\bibitem{zhou1} T. Zhou and J. X. Li, Phys. Rev. B {\bf 69}, 224514 (2004).
\bibitem{zhou2} T. Zhou and J. X. Li, Phys. Rev. B {\bf 72}, 134512 (2005).



\bibitem{kuro} K. Kuroki, S. Onari, R. Arita, H. Usui, Y. Tanaka,
H. Kontani, and H. Aoki, Phys. Rev. Lett. {\bf 101}, 087004 (2008).
\bibitem{maie} T. A. Maier, S. Graser, D. J. Scalapino, and P. Hirschfeld, Phys. Rev. B {\bf 79}, 134520 (2009).
\bibitem{kemp} A. F. Kemper, T. A. Maier, S. Graser, H. -P Cheng, P. J. Hirschfeld, and D. J. Scalapino, New J. Phys. {\bf 12}, 073030 (2010).
\bibitem{ygao} Y. Gao, T. Zhou, C. S. Ting, and W. P. Su, Phys. Rev. B {\bf 82}, 104520 (2010).
\bibitem{ygao1} Y. Gao, Y. Yu, T. Zhou, H. Huang, and Q. H. Wang, Phys. Rev. B {\bf 96}, 014515 (2017).
\bibitem{jianxin} J. X. Li and Z. D. Wang, Phys. Rev. B {\bf 70}, 212512 (2004).
\bibitem{chub} A. V. Chubukov and L. P. Gor'kov, Phys. Rev. Lett. {\bf 101}, 147004 (2008).
\bibitem{vign} B. Vignolle, S. M. Hayden, D. F. McMorrow, H. M. R$\mathrm{\o}$nnow, B.
Lake, C. D. Frost, and T. G. Perring, Nat. Phys. {\bf 3}, 163 (2007).
\end{thebibliography}
\end{document}